# Superior electrochemical performance of zinc-ion batteries with fine-grained and textured zinc anode produced by high-pressure torsion

Xinxin Hu[1], Shivam Dangwal[2,3], Xucheng Wang[1], Fan Zhang[1,*], Haijuan Kong[1], Jun Li[1], Kaveh Edalati[2,3,*]

[1] School of Materials Science and Engineering, Shanghai University of Engineering Science, Shanghai 201620, China
[2] WPI, International Institute for Carbon-Neutral Energy Research (WPI-I2CNER), Kyushu University, Fukuoka 819-0395, Japan
[3] Department of Automotive Science, Graduate School of Integrated Frontier Sciences, Kyushu University, Fukuoka, Japan

Zinc-ion batteries are promising alternatives to lithium-ion batteries, offering advantages in safety, cost, and environmental impact. However, their performance is often limited by the functioning of the zinc anode. This study employs severe plastic deformation via the high-pressure torsion (HPT) method to enhance the electrochemical performance of zinc anodes. HPT reduced the grain size from >1000 μm to 20 μm and introduced a (002) basal texture. The battery assembled with HPT-processed zinc demonstrated improved cycling stability, rate performance, and specific discharge capacity (>500 mAh/g at 0.5 A/g after 50 cycles), particularly at high current densities. This performance enhancement was attributed to grain-boundary and texture effects on improved ion transfer (confirmed by electrochemical impedance spectroscopy), fast redox reaction kinetics (confirmed by cyclic voltammetry), and reduced corrosion (confirmed by microscopy and potentiodynamic polarization test). This study highlights the potential of severely deformed materials with textured fine grains for advanced rechargeable battery technologies.

*Corresponding authors:
K. Edalati (E-mail: kaveh.edalati@kyudai.jp; Tel: 0081-92-802-6744)
F. Zhang (E-mail: fzhang@sues.edu.cn; Tel: +86-21-6779-1203)

# 1. Introduction

The global demand for efficient, sustainable and safe energy storage solutions is escalating due to the widespread adoption of electric vehicles, renewable energy systems and portable electronic devices [1]. While lithium-ion batteries have overpassed nickel-metal hydride batteries and dominated the market, concerns over lithium resource scarcity, safety and environmental impact have driven the search for alternative batteries [2]. Zinc-based batteries have emerged as a viable alternative due to the abundance of zinc, its low cost and its non-toxic nature [3]. Zinc-based batteries are classified into two main types: zinc-air batteries and zinc-ion batteries [3]. While zinc-air batteries generate power through the reaction of zinc with oxygen from the air [4], zinc-ion batteries operate through the intercalation and de-intercalation of zinc ions into a host material, typically in a closed system [5]. Unlike zinc-air batteries, which are limited by their reliance on atmospheric oxygen and associated challenges like $CO_2$ sensitivity, zinc-ion batteries offer more stable and controllable electrochemical processes, leading to better cycle life and higher power output [5].

Despite their advantages, zinc-ion batteries face several challenges with their cathode and anode materials that must be addressed. While various compositions were developed as active cathode materials [6,7], the flexibility in modifying the anode material is less because high-purity zinc is usually used [5]. One primary issue with the anode materials is the formation of zinc dendrites, which can lead to short-circuiting and capacity loss over time [8,9]. Additionally, the electrochemical performance of anodes can be hindered by the passivation and corrosion of zinc, which reduce ion transport efficiency, increase polarization and diminish overall battery performance [10]. Numerous studies attempt to enhance the cycling stability of the zinc-ion batteries by adding a secondary material to its electrodes. Polydopamine regulation [11], using metal-organic frameworks functionalized separators [12] or adding a fast ion conductor protective layer [13] are some examples of such attempts. One potential strategy to deal with these drawbacks without the use of an additional material is the modification of microstructure, which has been occasionally used for anode [14] and cathode [15] materials. The control of the zinc anode texture to (002) is another solution, because (002) is considered the most stable and corrosion-resistant atomic plane of zinc with relatively low surface energy [16,17].

Severe plastic deformation (SPD) methods offer promising solutions to modify the microstructure of materials by refining grain size and controlling texture [18,19]. High-pressure torsion (HPT), illustrated in Fig. 1a, is one of the most popular SPD techniques that simultaneously applies pressure and torsional shear strain to a disc-shaped material [20,21]. HPT was used in a few studies to process pure zinc and examine grain size [22,23], hardness [24], tensile properties [25] and biocompatibility [26]. However, there have been no attempts so far to use HPT-processed zinc in zinc-ion batteries, although a few studies used HPT for processing other battery materials [27-32].

This study aims to evaluate the performance of HPT-processed zinc anodes in zinc-ion batteries, focusing on their electrochemical properties and potential advantages over traditional zinc anodes. It is found that HPT refines the microstructure to equiaxed grains, introduces (002) texture, enhances the stability of the anode material, improves ion transport and redox reaction kinetics, and ultimately leads to better cycling stability and rate capability in zinc-ion batteries.

# 2. Experimental Methods
## 3.1 Preparation and Characterization of Zinc Anodes

Zinc discs, having a 10 mm diameter and 1 mm thickness with 99.99% purity, were received as anode materials in the annealed condition (523 K for 1 h). Some discs were kept in the annealed condition, while some others underwent HPT processing at room temperature, applying a pressure of 6 GPa with 10 revolutions at a rotation speed of 1 rpm, as attempted earlier [22-24]. The discs after annealing and HPT were polished on both upper and lower surfaces to adjust their thickness to 0.6 mm for use as the anode of zinc-ion batteries.

The discs were first characterized by X-ray diffraction (XRD) using a Cu Kα X-ray source. Moreover, their microstructure was examined by scanning electron microscopy (SEM). For SEM, discs were prepared by mechanical grinding using sandpapers, fine polishing using 9 and 3 µm diamond suspensions dispersed on buffs, and final polishing using 60 nm particle size colloidal silica suspension dispersed on a buff. SEM was conducted using an acceleration voltage of 15 kV by taking secondary electron images and electron backscatter diffraction (EBSD) mapping with step sizes of 10 µm and 3 µm for the annealed and HPT-processed samples, respectively. The analysis of EBSD data was done using the MTEX toolbox in MATLAB [33]. Grains were determined by taking a misorientation threshold of 2-15° for low-angle grain boundaries and >15 ° for high-angle grain boundaries. The grains smaller than 3 pixels in length were removed and merged into the surrounding pixel. The EBSD images were subjected to image processing by smoothing the grain boundaries and coloring unidentified small pixels with the average orientation of the surrounding grain using the object property meanOrientation [33]. The average grain size was calculated by considering both low- and high-angle grain boundaries.

## 2.2 Preparation and Testing of Batteries

For making batteries, with components shown in Fig. 1b, vanadium dioxide ($VO_2$) was selected as the cathode material due to its high theoretical capacity and ability to reversibly intercalate zinc ions [7,34]. The $VO_2$ powder was prepared through a hydrothermal synthesis process to achieve the desired crystalline phase. To do this, 600 mg of $V_2O_5$ and 800 mg of $H_2C_2O_4$ were mixed with 30 mL of deionized water and stirred at 333 K for 5 h. The solution was then subjected to hydrothermal reaction at 433 K for 12 h. The solid $VO_2$ phase was separated from the solution by centrifugation and dried under vacuum at 333 K for 12 h. The $VO_2$ powder was in the shape of nanorods with a monoclinic structure and the C2/m space group ($a$ = 12.030 Å, $b$ = 3.693 Å, $c$ = 6.420 Å, $α$ = 90º, $β$ = 106.6º and $γ$ = 90º), as shown in Fig. 2 using (a) SEM and (b) XRD. The cathode was fabricated by mixing $VO_2$ powder with acetylene black and polyvinylidene fluoride (PVDF) binder in a 7:2:1 weight ratio, dispersing the mixture in N-methyl-2-pyrrolidone (NMP), and casting the slurry onto a stainless steel mesh.

The zinc-ion batteries (Zn‖Zn symmetric and Zn‖$VO_2$) were assembled in a stainless steel CR2032 coin cell configuration under an air atmosphere, as attempted earlier [6,7,34]. Fig. 1b shows the different components of the batteries before and after assembling. For Zn‖Zn symmetric cells, HPT-processed or annealed zinc discs served as both the anode and cathode. For Zn‖$VO_2$ full cells, HPT-processed or annealed zinc discs served as the anode, with the $VO_2$-coated stainless steel mesh as the cathode. A glass fiber separator was used to prevent short circuits, and 3 M zinc trifluoromethanesulfonate ($Zn(CF_3SO_3)_2$) aqueous solution served as the electrolyte. The assembled cells were sealed and stabilized for 3 h before electrochemical testing. The electrochemical performance of the cells was evaluated using a potentiostat/galvanostat battery-testing system. The Zn‖Zn symmetric cells were subjected to galvanostatic charge-discharge cycling tests under a current density of 2 mA/cm$^2$ with a capacity of 0.5 mAh/cm$^2$ for determining the electrochemical performance of the HPT-processed and annealed zinc anodes. For Zn‖$VO_2$ full cells, galvanostatic charge-discharge tests were conducted at various current densities (0.2, 0.5, 1, 3 and 5 A/g) to assess the specific capacity, cycling stability and rate performance. Cyclic voltammetry and electrochemical impedance spectroscopy were also performed to analyze the redox behavior and charge transfer resistance of the batteries. The zinc anode materials after testing in the battery were examined by optical photography and SEM using an acceleration voltage of 3 kV.

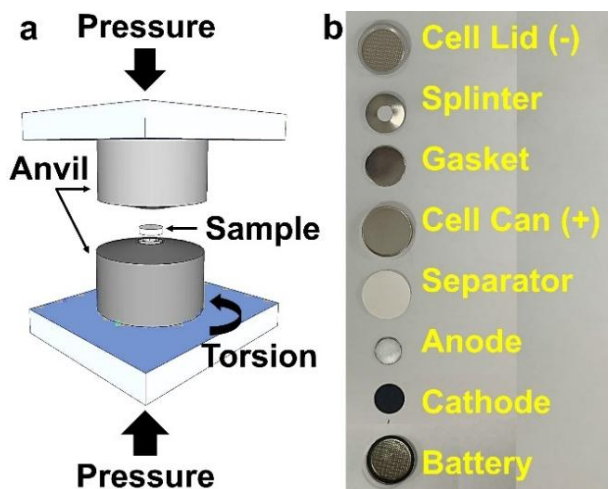

Figure 1. (a) Schematic representation of HPT method. (b) Components of zinc-ion batteries before and after assembling using HPT-processed zinc anode material. Components are assembled in the order of cell can, cathode, separator, anode, gasket, splinter, and cell lid (from the bottom to top).

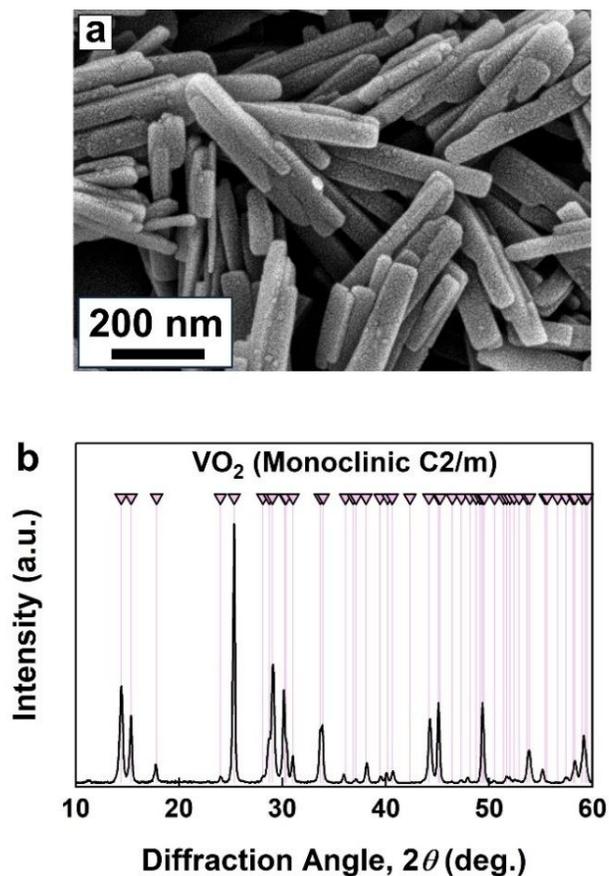

Figure 2. (a) SEM image and (b) XRD pattern for $VO_2$ powder used as cathode material.

**2.3 Corrosion Test**

Potentiodynamic polarization (PDP) tests were performed on both annealed and HPT-processed zinc anodes in an aqueous zinc trifluoromethanesulfonate solution using an electrochemical workstation with a typical three-electrode system. The zinc anode served as the working electrode, a platinum plate as the counter electrode, and Ag/AgCl as the reference electrode. The scan rate was set to 1 mV/s, and the potential range was -1.5 to 0.5 V (vs. Ag/AgCl). After the tests, the values of corrosion potential ($E_{corr}$) and corrosion current density ($i_{corr}$) were calculated from the Tafel zone of PDP curves.

## 3. Results
### 3.1 Microstructural Analysis

Structural analysis using XRD, as shown in Fig. 3, indicates that both annealed and HPT-processed samples have a hexagonal close-packed (HCP) structure consistent with PDF Card No. 03-065-3358. No phase transformation occurs by HPT processing, but the main change is increasing the peak intensity of the (002) plane suggesting a basal texture occurs. Since the main slip system in zinc is {001}<110>, an increase in the intensity of the basal plane peak due to HPT processing is expected, as this behavior has been reported in metals with an HCP crystal structure and a basal slip mechanism [18].

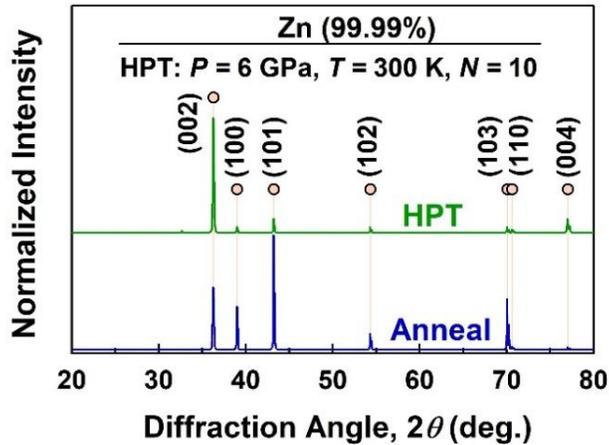

Figure 3. XRD profiles of zinc anode materials in the as-received annealed condition and after HPT processing.

SEM-EBSD was used to analyze the microstructure of the samples. As shown in Fig. 4a in the normal direction, grain sizes for the annealed sample are larger than 1000 μm. After HPT processing, as shown in Fig. 4b and 4c in the normal and transverse directions, respectively, the grain size is significantly reduced to 20 μm by HPT processing. Such small grain sizes after HPT processing, which are consistent with earlier publications [22,23,25], are expected to improve the resistance of zinc anode to dendrite formation and facilitate faster ion transport during cycling [14,15]. Another feature in the EBSD analysis of the HPT-processed sample, which is shown more clearly in the inverse pole figure counter of Fig. 4d, is the development of a basal texture that agrees well with XRD analysis. Such a texture is appropriate for anode materials of zinc-ion batteries because the (002) plane is the most stable and corrosion-resistant atomic plane of zinc [16,17].

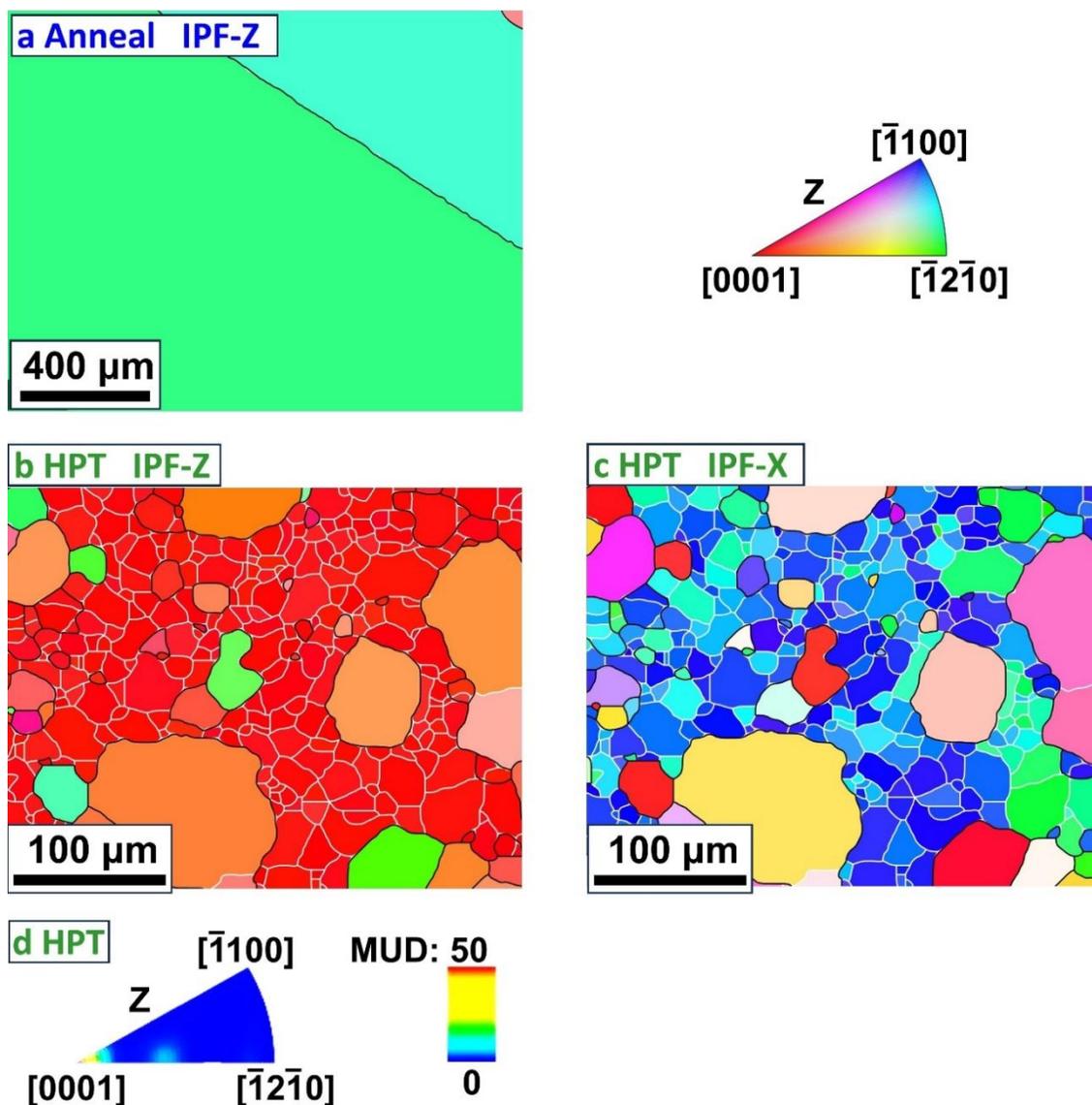

Figure 4. EBSD crystal orientation maps for (a) annealed zinc in Z (normal) direction, (b) HPT-processed zinc in Z direction and (c) HPT-processed zinc in X (transverse) direction, where white and black lines show low- and high-angle grain boundaries. (d) Inverse pole figure (IPF) counter for HPT-processed zinc in Z direction, where MUD stands for multiple of uniform density.

### 3.2 Electrochemical Performance

The Zn∥Zn symmetric cells were first examined by subjecting them to galvanostatic charge/discharge cycling tests to evaluate the stability of $Zn^{2+}$ plating/stripping under a current density of 2 mA/cm$^2$ with a capacity of 0.5 mAh/cm$^2$. As shown in Fig. 5, the voltage of both the batteries assembled using the annealed and HPT-processed materials fluctuates with increasing cycling time. However, the voltage of the battery assembled with annealed zinc drops drastically after cycling for 141 h (i.e. 242 cycles of charging/discharging), indicating the occurrence of a short circuit. In contrast, the voltage of the battery assembled with HPT-processed zinc drops after 337 h (i.e. 674 cycles) but still can fluctuate and finally short-circuit after cycling for 563 h (i.e. 1126 cycles of charging/discharging). It should be noted that the stability of these HPT-treated anodes can be further improved beyond 337 h in future works by using protecting layers on the surface, as attempted in earlier studies [35,36].

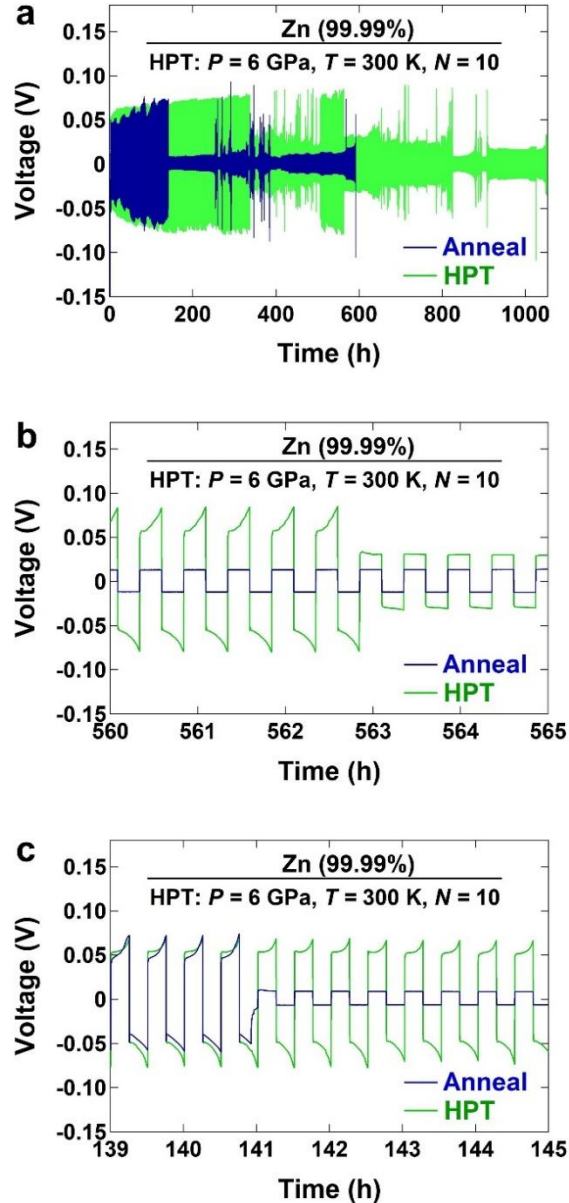

Figure 5. Galvanostatic charge/discharge cycling tests under a current density of 2 mA/cm$^2$ with a capacity of 0.5 mAh/cm$^2$ for Zn||Zn symmetric cells assembled using annealed and HPT-processed zinc anode materials, where (b) and (c) are magnified views of (a).

The electrochemical performances of Zn||VO$_2$ full batteries were also examined by cyclic test in a constant current density of 0.5 A/g and later by rate performance test by cycling the batteries at increasing current densities of 0.2, 0.5, 1, 3 and 5 A/g, as shown in Fig. 6a and 6b, respectively. The electrochemical performance of the zinc-ion batteries assembled with HPT-processed zinc anodes is significantly better than that of annealed zinc anodes. At a current density of 0.5 A/g, the HPT-processed anode retains a high discharge-specific capacity of 520 mAh/g after 50 cycles, representing only a slight decrease from its initial capacity of 550 mAh/g. This slight

decrease in capacity after 50 cycles suggests the need for long-term cycling tests in future studies. In contrast, the annealed zinc anode exhibits a rapid decline in capacity from 400 mAh/g to 220 mAh/g under the same conditions. In addition to cycling stability, the HPT-processed zinc anode exhibits excellent rate capability, maintaining discharge specific capacities of 582, 562, 540, 468 and 382 mAh/g at current densities of 0.2, 0.5, 1, 3 and 5 A/g, respectively. When the current density of HPT-processed zinc anode is reduced back to 0.2 A/g, the specific capacity recovers to 570 mAh/g, demonstrating a high capacity retention rate of 98.1%. In contrast, the annealed zinc anode displays lower specific capacities at each current density except for 0.2 A/g and a lower capacity retention rate of 88.9% when the current density is returned to 0.2 A/g. Here, it should be noted that the differences in capacities observed at a current density of 0.5 A/g in Fig. 5a and 5b are likely due to better activation of the anode material when it is first cycled at a lower current density (a lower current density results in less overpotential, as described by the Butler-Volmer equation) [37]. Nevertheless, these results underscore the advantages of HPT in enhancing the cycling stability and rate performance of zinc-ion batteries, features that are of significance in rechargeable batteries [1-5].

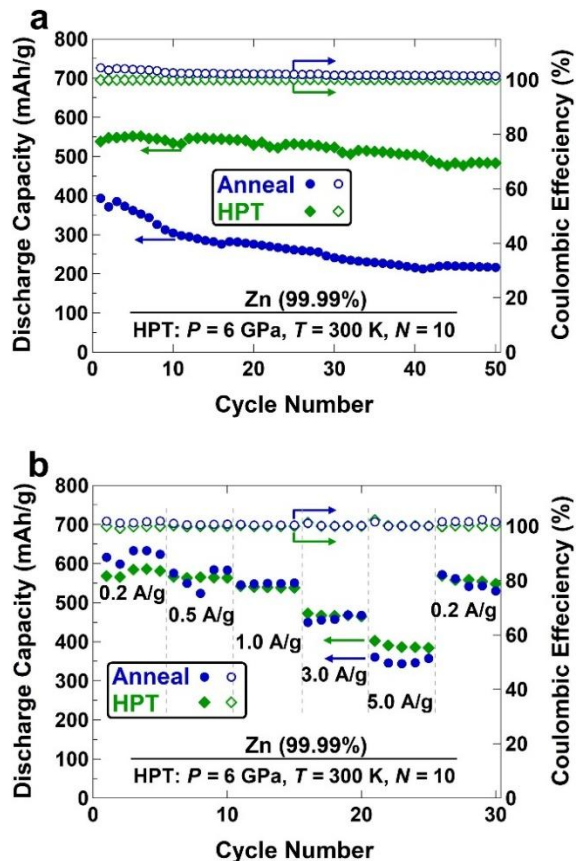

Figure 6. Specific discharge capacity versus cycle number in (a) cyclic test for a constant current density of 0.5 A/g and (b) rate performance test by cycling at increasing current densities of 0.2, 0.5, 1, 3 and 5 A/g for Zn||VO$_2$ full batteries assembled using annealed and HPT-processed zinc anode materials.

To understand the reasons for better performance of HPT-processed zinc anode, redox reactions and charge transfer resistance were examined by cyclic voltammetry and electrochemical impedance spectroscopy. As shown in Fig. 7a, the cyclic voltammetry curves of the HPT-processed zinc display sharp and well-defined redox peaks, indicating efficient and reversible zinc ion intercalation and de-intercalation [8-10]. In contrast, the annealed zinc exhibits broader peaks, suggesting slower kinetics and greater polarization. Electrochemical impedance spectrum analysis, shown in Fig. 7b, further confirms the lower charge transfer resistance of the HPT-processed anode. Semicircular curves, which are indicative of the charge transfer impedance of batteries, exhibit a lower radius for the HPT-processed anode. The electrochemical parameters including charge transfer resistance ($R_{ct}$) are listed in Table 1. The simulated $R_{ct}$ value for HPT-processed zinc (45.89 $\Omega \cdot cm^2$) is less than that of the annealed zinc (221.4 $\Omega \cdot cm^2$), indicating that its interface reaction with electrolyte is faster than the annealed zinc [8-10]. Moreover, the slope of the straight line that appeared in the low-frequency region of the electrochemical impedance spectrum is larger for the HPT-processed anode, indicating a faster ion diffusion. Enhanced redox reaction together with improved charge transfer for the HPT-processed electrode should contribute to better charge-discharge performance of the assembled battery [14-17].

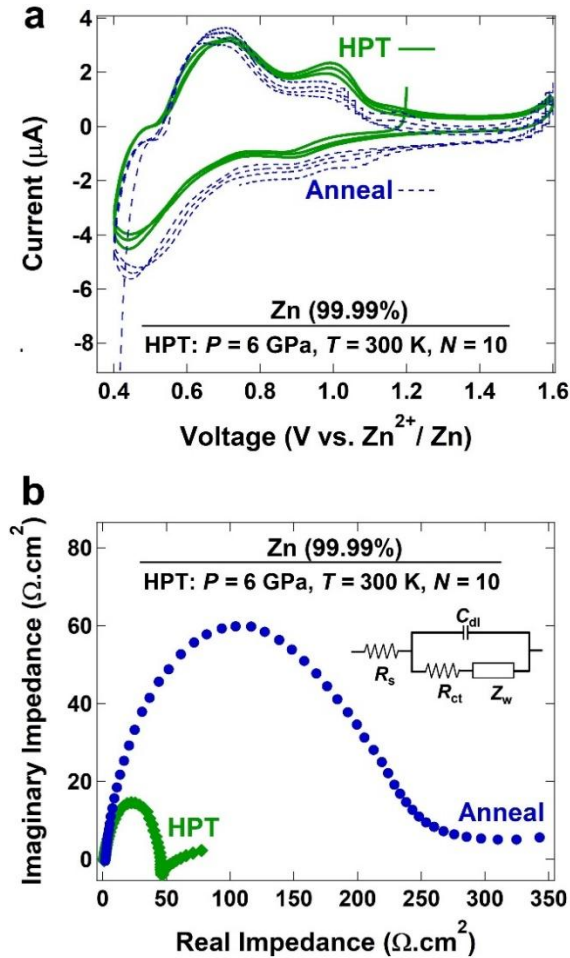

Figure 7. (a) initial three cyclic voltammetry curves and (b) electrochemical impedance spectra and equivalent circuit (insert image) for batteries assembled using annealed and HPT-processed zinc anode materials. Inset in (b): EIS equivalent circuit, where $R_s$ is solution resistance, $R_{ct}$ is charge transfer resistance, $C_{dl}$ is double layer capacitance and $Z_W$ is Warburg impedance.

Table 1. Simulated electrochemical parameters from EIS curves, where $R_s$ is solution resistance, $R_{ct}$ is charge transfer resistance, $C_{dl}$ is double layer capacitance, $Y_0$ is admittance parameter in constant phase element and n is constant phase element exponent.

| Sample | $R_s$ (Ω·cm$^2$) | $R_{ct}$ (Ω·cm$^2$) | $C_{dl}$ | |
|---|---|---|---|---|
| | | | $Y_0$ (S·cm$^{-2}$·s$^{\wedge}$n) | n |
| Anneal | 1.829 | 221.4 | 1.788 x 10$^{-4}$ | 0.7724 |
| HPT | 1.342 | 45.9 | 7.45 x 10$^{-5}$ | 0.8223 |

Fig. 8 shows the condition of the zinc anode after cycling for (a, c) annealed material and (b, d) HPT-processed material. While large localized corroded or mechanically damaged regions are formed in the annealed material, the size and area fractions of these regions are considerably smaller for the HPT-processed sample. A higher corrosion resistance of the HPT-processed sample can be understood from potentiodynamic polarization tests and calculated $E_{corr}$ and $i_{corr}$ values shown in Fig. 9. It is clear that HPT-processing slightly shifts $E_{corr}$ of zinc anode to a noble direction and results in a smaller $i_{corr}$ level, indicating a better corrosion resistance. This suggests that HPT chemically stabilizes the zinc anode, a fact that was also reported in other severely deformed materials used for different applications [18,19,38].

## 4. Discussion

This study demonstrates a new potential application for severely deformed materials which have received attention for different functionalities [18-21], including battery applications [27-32]. Here, a question is raised that needs further discussion: what is the mechanism of good performance of the battery assembled using HPT-processed zinc? The reduction of grain size in zinc anodes from >1000 µm to 20 µm and the development of basal texture through HPT can affect dendrite formation, ion transfer, redox reaction kinetics coupled with polarization, and mechanical/chemical stability.

*Dendrite formation:* Fine-grained zinc anodes with uniform microstructure is less prone to dendrite formation. The dendrite formation is a significant issue in zinc-based batteries that leads to short-circuiting and reducing battery life [14,39-41]. The formation of dendrites is particularly problematic at high current densities [39], while HPT leads to superior performance at high current densities. The likelihood of dendritic growth in fine-grained structures is less and thus HPT can diminish this problem by grain refinement [41]. Moreover, the basal texture developed by HPT also reduces the sensitivity to the dendrite formation [16,17,40].

*Ion transfer:* Smaller grain sizes increase the fraction of grain boundaries, which act as fast pathways for diffusion and thereby ion transport, improving the electrochemical performance of zinc-ion batteries [41,42]. The enhanced ion transport (confirmed by electrochemical impedance spectra in Fig. 7b) reduces the diffusion path lengths for zinc ions, facilitating quicker and more uniform ion distribution across the anode. This leads to a reduction in polarization during charge-discharge cycles, enhancing both the rate performance and cyclic stability [41,42]. This study shows that the HPT-processed zinc anode maintains a high specific capacity of 520 mAh/g after 50 cycles at a current density of 0.5 A/g, significantly outperforming the annealed zinc anodes due to a large fraction of grain boundaries with preferred orientation [42].

*Redox reaction kinetics and polarization*: Fine-grained structure can enhance the kinetics of redox reactions. An HPT-induced fine-grained structure increases the total fraction of grain boundaries as active sites for redox reactions [18,19], meaning more sites are available for zinc ion intercalation and de-intercalation. This abundance of active sites facilitates faster and more uniform redox reactions, as examined by the cyclic voltammetry analysis in Fig. 7a. Moreover, in a fine-grained structure, the travel distance of zinc ions within each grain is shorter than in larger grains. This reduction in diffusion path length indicates that zinc ions quickly and easily migrate to/from the grain boundaries during intercalation and de-intercalation [14,41,42]. This reduced diffusion path in fine-grained material tends to lower polarization, i.e. reduce the voltage loss caused by various resistances within the battery during charging and discharging. Additionally, a fine-grained structure promotes a more uniform distribution of zinc ions across the anode, while in larger grains, certain areas might become more heavily utilized than others, leading to localized ion distribution. All these parameters can contribute to enhanced redox kinetics observed in zinc-ion batteries assembled with HPT-processed anodes [3-5].

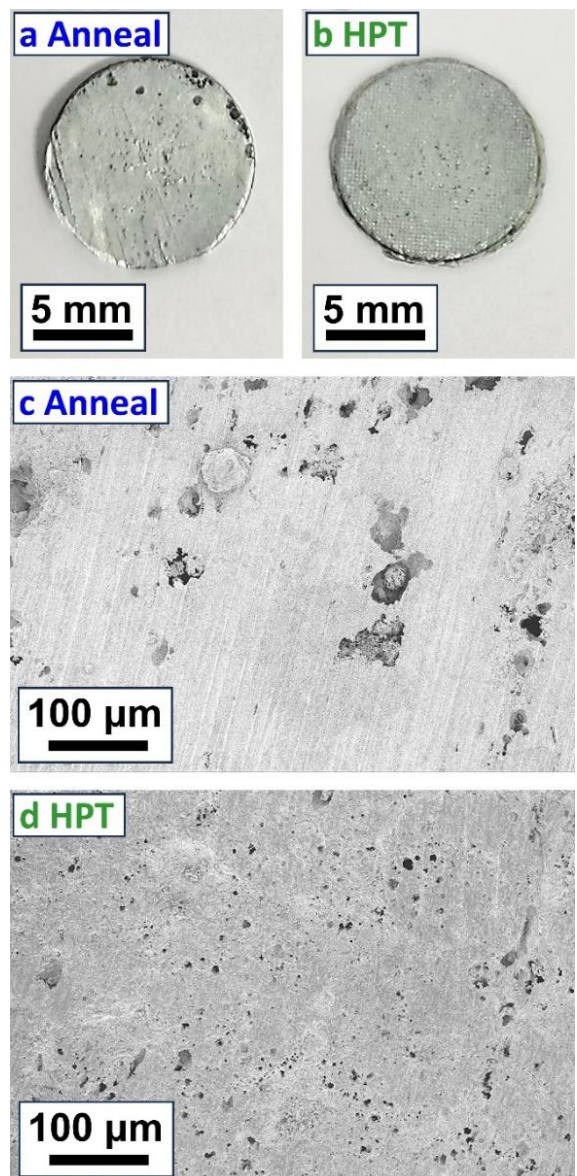

Figure 8. (a, b) Optical photographs and (b, d) SEM micrographs of (a, c) annealed and (b, d) HPT-processed zinc used as anode materials in zinc-ion batteries after 50 cycles of charge/discharge process.

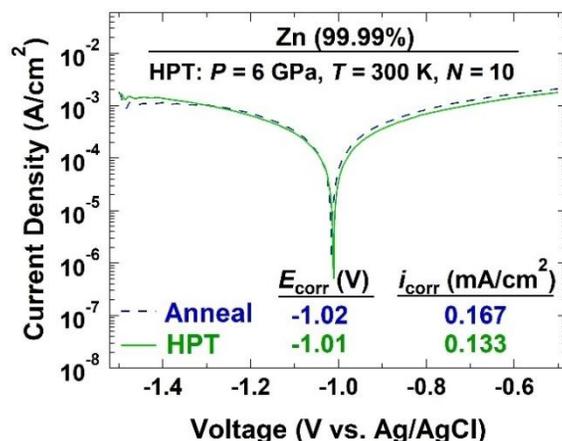

Figure 9. Potentiodynamic polarization curves and calculated corrosion potential ($E_{corr}$) and corrosion current density ($i_{corr}$) for annealed and HPT-processed zinc.

*Mechanical/chemical stability:* Besides the issues mentioned above, the authors consider that fine grains in severely deformed materials create a more mechanically stable structure that can be less prone to cracking and degradation during the repeated volume changes that occur during cycling tests [18,19]. Such structures are less prone to localized corrosion, an issue that was confirmed by examination of zinc anodes after the cyclic test of batteries in Fig. 8 and by corrosion tests in Fig. 9. Moreover, the basal texture developed by HPT can be another reason for less extent of localized corrosion because the (002) atomic plane is the most stable and corrosion-resistant surface of zinc due to its low surface energy [16,17].

## 5. Conclusions

In conclusion, this study demonstrates the significant impact of HPT on the performance of zinc-ion batteries. The HPT-processed zinc anode exhibits remarkable improvements in cycling stability, specific capacity and rate capability compared to the traditionally annealed zinc anode. The enhanced electrochemical performance is attributed to the refined microstructure and basal texture of the HPT-processed zinc, which subsequently reduces dendrite formation, improves ion transport, enhances the redox reactions and improves the mechanical/chemical stability. These findings highlight the potential of SPD techniques in advancing the performance of zinc-ion batteries, making them more competitive with other battery technologies.

## CRediT Authorship Contribution Statement

All authors: Conceptualization, Methodology, Investigation, Validation, Writing – review & editing.

## Declaration of competing interest

The authors declare no competing interests that could have influenced the results reported in the current article.

## Acknowledgments

This work is supported partly by Class III Peak Discipline of Shanghai - Materials Science and Engineering (High-Energy Beam Intelligent Processing and Green Manufacturing), and partly by the Japan Society for the Promotion of Science (JSPS) through a grant number JP22K18737.

**Data Availability**

The data will be made available upon request from the corresponding author.